\documentclass[a4paper,11pt]{article}
\pdfoutput=1 

\usepackage{epsfig}
\usepackage{subcaption}
\usepackage{graphicx}
\usepackage{soul}
\usepackage[english]{babel}
\usepackage{mathtools}
\usepackage{amssymb}
\usepackage{amsmath}    
\usepackage{jheppub}
\usepackage{braket}
\usepackage[normalem]{ulem}

\def	\be	{\begin{equation}}
\def	\ee	{\end{equation}}
\def	\nn	{\nonumber}

\title{Back(reaction) to the Future in the Unruh-de Sitter State}

\author[a]{Lars Aalsma,}
\author[b]{Maulik Parikh,}
\author[a]{and Jan Pieter van der Schaar}
\affiliation[a]{Institute for Theoretical Physics, Delta Institute for Theoretical Physics, University of Amsterdam, PO Box 94485, 1090 GL Amsterdam, The Netherlands}
\affiliation[b]{Department of Physics and Beyond: Center for Fundamental Concepts in Science\\ Arizona State University, Tempe, Arizona 85287, USA}
\emailAdd{l.aalsma@uva.nl}
\emailAdd{maulik.parikh@asu.edu}
\emailAdd{j.p.vanderschaar@uva.nl}

\abstract{Motivated by black hole physics, we define the Unruh state for a scalar field in de Sitter space. Like the Bunch-Davies state, the Unruh-de Sitter state appears thermal to a static observer. However, it breaks some of the symmetries of de Sitter space. We calculate the expectation value of the energy-momentum tensor in the Unruh-de Sitter state in two dimensions and find a non-vanishing flux of outgoing negative energy. Extrapolating the result to four dimensions, we argue that this backreacts on the initial de Sitter geometry semi-classically. Notably, we estimate that de Sitter space is destabilized on a timescale set by the gravitational entropy; analogous to black hole evaporation, the endpoint of this instability is a singular geometry outside the regime of effective field theory. Finally, we suggest that the Unruh-de Sitter state may be a natural initial state for patches of de Sitter space, and discuss the implications for slow-roll and eternal inflation, and for de Sitter thermodynamics.}

\begin{document} 
\maketitle
\flushbottom

\section{Introduction}
\label{sec:intro}

There have been contradictory claims about the quantum instability of de Sitter space. In much of the literature, a common and very reasonable approach is to start out with perturbative quantum fluctuations in the Bunch-Davies state. Since that state is invariant under the de Sitter isometry group, the question boils down to whether or not (and how) de Sitter symmetry becomes anomalous. Over the years different authors, employing a variety of techniques and methods, have reached different conclusions on the ultimate fate of de Sitter space. The predictions include exact all-order stability \cite{Rogers:1991qe,Busch:1992jw,Marolf:2010nz}, slow or fast decay \cite{Ford:1984hs,Ford:1985qh,Traschen:1986tn,Redmount:1988pg,Redmount:1989fm,Tsamis:1992sx,Anderson:2000wx,Brandenberger:2002sk,Polyakov:2007mm,Larjo:2011uh,Marozzi:2012tp,Anderson:2013zia,Dabholkar:2015qhk,Rajaraman:2016nvv,Markkanen:2017abw,Brandenberger:2018fdd,Saitou:2019jez}, and evolution towards a phase in which perturbative quantum field theory breaks down \cite{Dvali:2014gua}. The different approaches, whose consistency and interpretation is not always transparent, make it difficult to compare results and reach consensus on this important question. This is particularly confusing in light of the fact that one would expect to be able to make unambiguous statements, at least in perturbation theory, as long as the de Sitter curvature is small, since an effective field theory description should be sufficiently accurate then.   

Here we will propose yet another approach to the question of the quantum stability of de Sitter space. Our approach is motivated by lessons learned from the evaporation of black holes. Recall how we know that nonextremal black holes are quantum mechanically unstable. The emission of Hawking radiation from the horizon into the asymptotically flat geometry, where energy is well-defined and conserved, is not by itself sufficient to conclude that a black hole must evaporate: it is also necessary that there not be any compensating incoming energy flux from infinity. Indeed, for black holes there are two special quantum states which have Hawking radiation: the Unruh \cite{Unruh:1976db} and the Hartle-Hawking state \cite{Hartle:1976tp}. Although an asymptotic observer measures thermal radiation in both of these states, it is only in the Unruh state that the black hole decays. In the Hartle-Hawking state, boundary conditions are such that the outgoing flux of energy is balanced by incoming flux from infinity; this state is therefore typically used to describe the maximally-extended eternal black hole in thermal equilibrium, rather than the evaporating black hole. In contrast, in the Unruh state there is no incoming flux from infinity and the outgoing flux is no longer compensated for: the black hole evaporates. Note also that, whereas the vacuum expectation value of the energy-momentum tensor is regular everywhere in the Hartle-Hawking state, in the Unruh state it features a singularity on the past horizon. Despite this, the Unruh state is accepted as a physically viable state because it is regular on the future horizon; the putative singularity on the past horizon is occluded by the collapsing matter forming the black hole.

We will translate these observations to the context of de Sitter space. The de Sitter counterpart of the Hartle-Hawking state is the Bunch-Davies state, a state annihilated by the full $O(1,D)$ group of de Sitter isometries. Here we propose that one can also consider a de Sitter analogue of the Unruh state. As in the black hole case, this state imposes different boundary conditions for incoming and outgoing fluxes. It also breaks homogeneity by identifying a special static patch region. Correspondingly, the vacuum expectation value of the energy-momentum tensor is regular on the future horizon, but diverges on the past horizon. Nevertheless, as we will argue, this state is consistent and even reasonable from a certain point of view \cite{Greene:2005wk}. The first goal of this paper then is to explicitly construct the Unruh-de Sitter state.

The absence of an asymptotically flat region in de Sitter space means that one cannot rely on global energy conservation arguments to determine the backreaction in a given state \cite{Parikh:2004rh}; instead, a careful calculation of the expectation value of the energy-momentum tensor is required to analyze the stability of the background. Our second goal, then, is to compute the vacuum expectation value of the energy-momentum tensor in the Unruh-de Sitter state. Unfortunately, as in black hole geometries, a direct computation of this quantity is not straightforward \cite{Candelas:1980}. We therefore focus on 1+1 dimensional de Sitter space, where the calculation is more tractable and can be readily interpreted for the different states. For the Bunch-Davies state, we find the well-known result that the vacuum expectation value of the energy-momentum tensor is proportional to the metric, as it must be in order to preserve the de Sitter isometries. This renormalizes the bare cosmological constant but does not destabilize de Sitter space. But for the Unruh-de Sitter state, we find a quantum violation of the null energy condition, as with Hawking radiation from black holes.

We also find that the energy-momentum tensor is singular on the past horizon. Under certain assumptions, which we spell out, the 1+1 results can then be generalized to the s-wave sector of 3+1 dimensional de Sitter space. 

We then estimate the backreaction on the geometry. Although the expectation value of the energy-momentum tensor in the Unruh-de Sitter state is inhomogeneous, we can, to first approximation, analyze the instability at the level of the Friedmann equation. As happens with evaporating black holes, the direction of the instability in the Unruh state is not towards Minkowski space but rather towards a singular geometry whose description lies outside the semi-classical regime. Again in parallel with the black hole case, the time-scale of the instability is set by the gravitational entropy, that is $H \, t \sim M_p^2/H^2$. As such, the instability cannot explain the smallness of the cosmological constant (being both too slow and in the wrong direction) but when applied in the context of eternal inflation it could potentially destabilize Hubble-sized regions of de Sitter space.

This paper is organized as follows. In section \ref{sec:Rindlerspace}, we recall some elementary results about quantum fields in Rindler and Schwarzschild backgrounds. In particular, we review the different choices of state one can make and present the resulting vacuum expectation value of the energy-momentum tensor in those states. In section \ref{sec:dSspace}, we translate the problem to de Sitter space, defining the Unruh-de Sitter state. In section \ref{sec:emt}, we calculate the vacuum expectation value of the energy-momentum tensor in two and four dimensions. In section \ref{sec:backreaction}, we use the expectation value to analyze the instability of de Sitter space in the Unruh-de Sitter state. We give a bound on the maximal number of e-folds before the semi-classical description breaks down. For most of the paper we emphasize the mathematical consistency of the Unruh-de Sitter state, but, in section \ref{sec:conclusions} we argue that in many circumstances for which de Sitter space is used to approximate primordial or late-time cosmology, the Unruh-de Sitter state is a natural choice. We consider the potential implications of our results for (eternal) inflation and for the de Sitter swampland conjectures \cite{Obied:2018sgi,Ooguri:2018wrx}.

\section{States and backreaction in Rindler and Schwarzschild geometries} \label{sec:Rindlerspace}

\subsection{Rindler space in 1+1 dimensions}
Let us begin with a brief review of two-dimensional Rindler space. In Rindler coordinates the line element reads
\be
ds^2 = e^{2a\xi}\left(-d\tau^2 + d\xi^2\right) ~.
\ee
The Minkowski inertial coordinates $t$ and $x$ are related to Rindler coordinates in the right Rindler wedge, $x > |t|$, through
\begin{align}
t = \frac1{a}e^{a\xi}\sinh(a\tau)  \\
x = \frac1{a}e^{a\xi}\cosh(a\tau) \nn
\end{align}
Define lightcone coordinates with respect to Minkowski time $t$ and Rindler time $\tau$:
\begin{align}
U &= t-x ~, \qquad u=\tau-\xi ~.  \\
V &= t+x ~, \qquad v=\tau+\xi ~. \nn
\end{align}
They are related by
\begin{align}
U &= -\frac1a e^{-au}  \\
V &=+ \frac1a e^{av} \nn
\end{align}
in the right Rindler wedge, $U<0<V$. Then the line element becomes
\be
ds^2 = -dUdV = -e^{a(v-u)}dudv	 \label{uvRindlermetric}
\ee
Now consider a massless, minimally-coupled (therefore also conformal) scalar field. We can expand the field in Minkowski modes, for which positive frequency is defined with respect to $i \partial_t$, where $t$ is the proper time of an inertial observer. Upon quantization, we call the operators that multiply these modes the $\hat{a}$ operators. Alternatively, we can decompose the field in modes that have positive frequency with respect to $i \partial_\tau$, the proper time of a Rindler observer. We call the corresponding operators the $\hat{b}$ operators. For the reader's convenience the relevant details of quantum field theory in Rindler space are reviewed in the appendix.

For a conformal theory solutions to the wave equation split into left- and right-moving parts:
\be
\partial_U \partial_V \varphi = 0 \quad \Rightarrow \quad \varphi(U,V) = \varphi_U(U) + \varphi_V(V) ~.
\ee
There are then four special cases of vacuum states because we can select either the $a$ vacuum or the $b$ vacuum for the left-moving and right-moving modes independently. Choosing both left- and right-movers to be in the $a$ vacuum selects the Poincar\'e-invariant or Minkowski vacuum, $|0_M\rangle$. Choosing both to be in the $b$ vacuum selects the Rindler vacuum, $|0_R \rangle$. But note that it is also possible to make asymmetric choices, choosing an $a$ vacuum for the left-movers and a $b$ vacuum for the right-movers, or vice versa.

The vacuum expectation value of the energy-momentum tensor, which classically is given by
\be
T_{\mu \nu} = \partial_\mu \varphi \partial_\nu \varphi- \frac{1}{2} (\partial \varphi)^2 \eta_{\mu \nu} ~,
\ee
can be computed by expanding $\varphi$ in modes that solve the wave equation and evaluating the resulting mode sums. As we review in the appendix, this computation yields the well-known result
\begin{align}
\langle 0_R | T_{UU} | 0_R \rangle  - \langle 0_M | T_{UU} | 0_M \rangle &= - \frac1{48\pi U^2} ~, \nn \\
\langle 0_R | T_{VV} | 0_R \rangle  - \langle 0_M | T_{VV} | 0_M \rangle &= - \frac1{48\pi V^2} ~.
\end{align}
We see that, relative to the Minkowski vacuum, the Rindler vacuum has a negative energy density which is singular at the past ($V=0$) and future ($U=0$) acceleration horizons. These singularities make the Rindler vacuum unphysical. Note also that in principle one could make a hybrid, asymmetric choice to allow for states that are singular at only one of the horizons. We do not usually make such asymmetric vacuum choices in Minkowski space because even a singularity at one horizon would render it unphysical.

An entirely analogous story applies to quantum fields in the background of black holes, which we take here to be a Schwarzschild black hole for simplicity. The counterpart of the Poincar\'e-invariant state is the Hartle-Hawking state, while the counterpart of the Rindler vacuum is the Boulware vacuum. But in addition, for physical black holes we also have the Unruh state. This is an asymmetric state in which the outgoing ($U$) vacuum is chosen to be annihilated by the $\hat a$ operators, while the ingoing ($V$) vacuum is chosen to be annihilated by the $\hat b$ operators. This state is singular on the past horizon but regular everywhere else. In the context of a black hole that forms from a collapsing star, however, the Schwarzschild geometry is replaced at early times by the geometry of the collapsing star. This covers up the past horizon so that the would-be singularity there is of no physical relevance. Thus, the Unruh state is acceptable and in fact natural. As we shall see, the de Sitter counterpart of the black hole Unruh state is well-defined on an entire planar patch and might even be a natural alternative to the commonly used Bunch-Davies state. These results are summarized in Table 1.
\begin{table}[h]
\label{Table1}
\resizebox{\textwidth}{!}
{\begin{tabular} {c  c | c | c | c | c}
\hline
$U$ & $V$ & Name & Thermal flux & $T_{UU}+T_{VV}$ & $\langle T_{\mu\nu} \rangle$ singularities\\
\hline
$\hat a$ & $\hat a$ & Poincare/Hartle-Hawking & ingoing, outgoing & 0 & none \\
$\hat b$ & $\hat b$ & Rindler/Boulware	& none & - $\frac{1}{48 \pi} \left (\frac{1}{U^2} + \frac{1}{V^2} \right )$ & future + past horizon \\
$\hat a$ & $\hat b$ & -/Unruh	&	outgoing & - $\frac{1}{48 \pi} \frac{1}{V^2}$  & past horizon \\
$\hat b$ & $\hat a$ & -/Unruh' &	ingoing & - $\frac{1}{48 \pi} \frac{1}{U^2}$  & future horizon \\
\hline
\end{tabular}}
\caption{Different vacua in two-dimensional Minkowski and Schwarzschild spacetimes and their respective fluxes, null energies, and singularities.}
\end{table}
We have included the time-reversed Unruh$^\prime$ state, for which the singularity occurs on the black hole future horizon; this, of course, is not a physically reasonable state if one assumes that the equivalence principle holds and that nothing special therefore happens at the future event horizon. 

\section{The Unruh-de Sitter state} \label{sec:dSspace}

\begin{figure}[ht]
    \centering
    \includegraphics[scale=1.0]{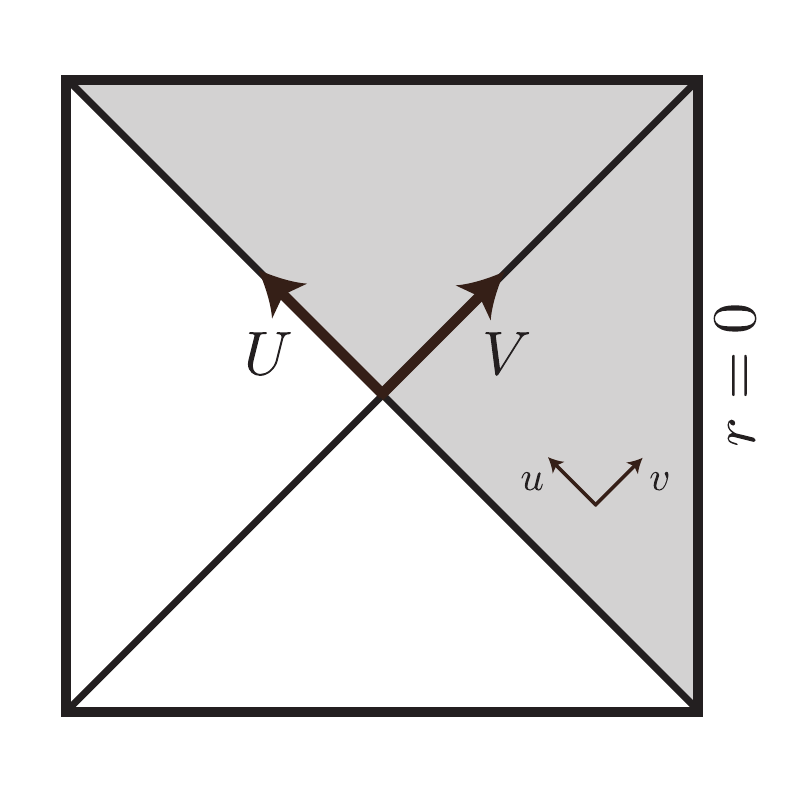}
    \caption{The de Sitter Penrose diagram, including the static ($u,v$) and global ($U,V$) lightcone coordinates, as defined. The planar patch is shaded gray.}
    \label{dScoordinates}
\end{figure}

Let us now define the Unruh state of a massless, minimally coupled scalar field in de Sitter space. For simplicity, we will first consider a scalar field in two-dimensional de Sitter space and later extend our result to four dimensions. Recall the line element in static coordinates
\be
ds^2 = -\left(1-H^2r^2\right)dt^2 + (1-H^2r^2)^{-1}dr^2~.
\ee
These coordinates cover only the static patch of de Sitter space. To cover the entire static patch region in two dimensions, the coordinate $r$ has to run over negative values as well. To be able to connect to the spherically symmetric four-dimensional case we need to perform a $\mathbb{Z}_2$ orbifold identifying $r$ and $-r$. This of course introduces a (mild) singularity at the origin, which will be important to keep in mind when discussing the two-dimensional energy-momentum tensor. For now we will just restrict to $r\geq 0$. In terms of the tortoise coordinate
\be
r_* = \frac1{2H}\log\left(\frac{1+Hr}{1-Hr}\right) ~,
\ee
one finds that two-dimensional de Sitter space is conformally flat
\be
ds^2 = (1-\text{tanh}^2(Hr_*))(-dt^2+dr_*^2) ~.
\ee
Following the conventions of Gibbons and Hawking \cite{Gibbons:1977}, we define the lightcone coordinates
\be
u = t+r_* ~, \qquad v = t - r_* ~,
\ee
in terms of which the positive frequency solutions to the wave equation are the same as in flat space:
\be \label{eq:staticmodes}
\varphi(u,v) =  \frac1{\sqrt{4\pi\omega}} \left( e^{-i\omega u} + e^{-i\omega v} \right)~.
\ee
We will also make use of planar coordinates for which the line element is
\be \label{eq:planar}
ds^2 = - d\tau^2 + e^{2H\tau} d\rho^2 = \frac{1}{(H \eta)^2} (- d\eta^2 + d\rho^2) 
\ee
The time coordinate $\tau$ is not to be confused with the Rindler time appearing earlier. Finally, we define global lightcone coordinates (see Figure \ref{dScoordinates})
which can be analytically continued to provide a global cover of de Sitter space. In terms of static coordinates, the global lightcone coordinates are
\be
U = -\frac{e^{-Ht}}{H}\sqrt{\frac{1-Hr}{1+Hr}} ~, \qquad V = \frac{e^{+Ht}}{H}\sqrt{\frac{1-Hr}{1+Hr}} ~.
\ee
In terms of planar coordinates the lightcone coordinates are
\be
U = \rho+\eta ~, \qquad V = \frac1{H^2(\rho-\eta)} ~,
\ee
where the static radial coordinate is not to be confused with the planar comoving radius. The line element in these coordinates is
\be
ds^2 = -\frac{4}{(H^2 UV-1)^2}dUdV ~.
\ee

Using these coordinates, the wave equation has positive frequency solutions given by
\be \label{eq:globalmodes}
\varphi(U,V) = \frac 1{\sqrt{4\pi\omega}} \left(e^{-i\omega U} + e^{-i\omega V} \right) ~.
\ee
The Bunch-Davies state, understood as the analogue of the Hartle-Hawking vacuum for black holes, is now defined as the state that is annihilated by the $\hat a$ operators that multiply the positive frequency modes \eqref{eq:globalmodes} for both incoming and outgoing modes, that is\footnote{The Bunch-Davies state is commonly defined by using mode functions that are positive frequency with respect to the time coordinate, $\eta$ in the limit $\eta\to-\infty$, used in the planar slicing of de Sitter space. This is equivalent to the state defined in (3.11) because at the past horizon $V=0$ the positive-frequency mode function $e^{-i \omega U}$ is also positive frequency with respect to $\eta$, as can be seen by expressing it in planar coordinates.}
\be \label{eq:defBD}
\hat a_{\rm in} | 0_{BD}\rangle = \hat a_{\rm out} | 0_{BD}\rangle = 0 ~.
\ee
The static vacuum, which is the counterpart of the Boulware state for black holes, can be defined as the state that is annihilated by the $\hat b$ operators that multiply the positive frequency modes \eqref{eq:staticmodes} for both incoming and outgoing modes:
\be
\hat b_{\rm in} | 0_S\rangle = \hat b_{\rm out} | 0_S\rangle = 0 ~.
\ee
Finally, the Unruh-de Sitter state is the state whose incoming modes (moving towards an $r=0$ observer from the past horizon) are annihilated by the $\hat a$ operator, but whose outgoing modes (moving away from an $r=0$ observer towards the future horizon) are annihilated by the $\hat b$ operator:
\be
\hat a_{\rm in} | 0_U\rangle = \hat b_{\rm out} | 0_U\rangle = 0 ~.
\ee
That is, incoming modes are positive frequency with respect to the canonical affine parameter on the past horizon, $U$, while outgoing modes are positive frequency with respect to the static time coordinate, $t$. It is this state that we would like to study in more detail.

Now, in order to define the Unruh state, we made use of the fact that because of the conformal nature of the field in two dimensions, the left-moving and right-moving sectors decoupled and their vacua could be defined independently. In four dimensions, the incoming and outgoing sectors do not decouple. However, we can still define their vacua independently. To see this, note that the wave equation for a (possibly massive) scalar field expressed in static coordinates is
\be
0 = (\Box - m^2) \varphi = \left ( - \partial_t^2 + \partial_{r_*}^2 + (1-H^2 r^2) \left [ \frac{2}{r} \partial_{r_*} + \frac{1}{r^2} \nabla^2_{S^2} - m^2 \right ] \right ) \varphi
\ee
Near the horizon, this reduces to $(- \partial_t^2 + \partial_{r_*}^2 ) \varphi = 0$ which can indeed be written in terms of decoupled incoming and outgoing modes. The operator coefficients of those modes can now be used to define the $U$ and $V$ vacua independently.

The Unruh-de Sitter state is essentially a hybrid of the Bunch-Davies and the static de Sitter vacuum, making an asymmetric state choice for incoming (Bunch-Davies) and outgoing (static) modes. As a consequence, it breaks the de Sitter symmetries in the outgoing sector, by explicitly identifying a special free-falling observer, as in the static vacuum. It follows that the Unruh-de Sitter state preserves the same symmetries as the static vacuum, namely $SO(3) \times R$. This will have important consequences for the analysis of the expectation value of the energy-momentum tensor and the corresponding consistency and stability of the state, to which we now turn.  

\section{Expectation value of the energy-momentum tensor} \label{sec:emt}
To evaluate the energy-momentum tensor in any particular state, one could expand the field $\varphi$ using the corresponding modes in the different sectors, insert that into
\be
\langle T_{\mu \nu} \rangle = \langle \partial_\mu \varphi \partial_\nu \varphi \rangle - \frac{1}{2} g_{\mu\nu}\langle\partial_\rho\varphi\partial^\rho\varphi\rangle ~,
\ee
and calculate by brute force. The disadvantage of this direct procedure, however, is that a proper regularization scheme is required to obtain a finite answer, which is of course somewhat subtle in curved spacetimes \cite{Birrell:1982ix}. Here we will bypass the subtleties by employing an approach pioneered by Christensen and Fulling \cite{Christensen:1977jc}, which was applied to two-dimensional de Sitter space not much later \cite{Lohiya_1978}. We will impose various consistency conditions that the components of $\langle T_{\mu\nu} \rangle$ should satisfy, independent of the state under consideration. This allows us to construct the expectation value of the energy-momentum tensor without the need for an explicit regularization procedure that might obscure the relevant physics.

\subsection{Consistency conditions in two dimensions}

The vacuum states that we would like to consider are all spherically symmetric, and time-translation invariant, but not necessarily homogeneous. Hence will allow the energy-momentum tensor components to depend only on the radius: $\langle T_{\mu\nu} \rangle = \langle T_{\mu\nu} (r) \rangle$. As we shall see, this will be all the freedom needed to describe a two-parameter family of states that includes all the states we are interested in.

Now, using static coordinates in two dimensions, the conservation of the energy-momentum tensor
\be
\nabla_\mu \langle T^{\mu\nu} \rangle = 0 ~,
\ee
leads to two differential equations for the components of the energy-momentum tensor:
\begin{align}
\partial_r \langle T^{tr} \rangle &= \frac{2H^2r}{1-H^2r^2} \langle T^{tr} \rangle ~,\\
\partial_r \langle T^{rr} \rangle &= -H^2r \langle T^\mu_{\,\,\,\mu} \rangle  ~. \nn
\end{align}
The first equation is solved by
\be
\langle T^{tr} \rangle = \frac{\Phi}{1-H^2r^2} ~,
\ee
where $\Phi$ is a constant. A non-zero off-diagonal component of the energy-momentum tensor implies that there is a net energy flux; positive $\Phi$ corresponds to positive energy in the outgoing direction. Note that the incoming or outgoing direction is uniquely defined only when restricting to $r\geq0$. Crossing from positive to negative $r$, requiring just incoming or outgoing flux implies the constant $\Phi$ has to change sign, producing a discontinuity at the origin related to an orbifold singularity. This subtlety is absent in the solution of the differential equation for $T^{rr}$, where we use the fact that a massless scalar field in two dimensions is conformally invariant. As a consequence the trace of the energy-momentum tensor picks up a state-independent conformal anomaly. For two-dimensional de Sitter space the conformal anomaly is \cite{Candelas:1978gf}
\be
\langle T^\mu_{\,\,\,\mu} \rangle = \frac{H^2}{12\pi} ~.
\ee
We then find
\begin{align}
\langle T_{tt} \rangle &= \frac{H^2}{24\pi}(H^2r^2-2) + \Omega ~, \\
\langle T_{rr} \rangle &=\frac1{(1-H^2r^2)^2}\left( -\frac{H^4r^2}{24\pi} + \Omega \right) ~. \nn
\end{align}
We see that the constant $\Omega$ is associated with the energy density measured by an observer at $r=0$. Together with the flux $\Phi$, these two constants parameterize the conserved energy-momentum tensor of any two-dimensional conformally-invariant scalar in a time-independent spherically-symmetric quantum state. 

Transforming to global lightcone coordinates, the result is particularly elegant:
\begin{align}
\langle T_{UU} \rangle &= -\frac{1}{48\pi U^2} + \frac{\Omega-\Phi}{2H^2U^2} ~, \\
\langle T_{VV} \rangle &= -\frac{1}{48\pi V^2} + \frac{\Omega+\Phi}{2H^2V^2} ~, \nn \\
\langle T_{UV} \rangle &= -\frac{H^2}{12\pi(H^2UV-1)^2} ~. \nn
\end{align}
The interpretation of the different components is as follows. The off-diagonal $\langle T_{UV} \rangle$ term is completely fixed by the conformal anomaly and is state-independent. The diagonal components describe the (null) energy present in the state under consideration. The $\langle T_{UU} \rangle$ component describes the energy density at a constant $U$ slice and is determined by the state selected for the incoming modes. Similarly, $\langle T_{VV} \rangle$ captures the energy along a constant $V$ slice and is fixed by the state selected for the outgoing modes. Note that for generic $\Omega$ and $\Phi$ parameters the energy-momentum tensor is singular at both the future $(U=0)$ and past horizon $(V=0)$. Only for special values of these parameters are the singularities absent. 

Different states can now be identified with particular values of the flux $\Phi$ and energy density parameter $\Omega$, or equivalently by specifying incoming and outgoing null energy. Notice that if we set the net flux $\Phi$ to zero, the $UU$ and $VV$ components take the same form. On the other hand, when $\Phi$ is non-vanishing, the additional contribution to the diagonal null components of the energy-momentum tensor carries different signs; a net incoming flux of positive energy particles, corresponding to negative $\Phi$, contributes positively to $\langle T_{UU} \rangle$ and negatively to $\langle T_{VV} \rangle$.

Since two-dimensional spacetimes are conformally flat, we can relate states constructed in Minkowski space to states in de Sitter space simply by performing a conformal transformation \cite{Candelas:1978gf}. Doing this for the Rindler vacuum, we obtain the static vacuum state in de Sitter space. The static vacuum is the empty state for a static observer at $r=0$. Hence, we anticipate that there should be no net flux and no additional energy density associated to particles in the static vacuum, suggesting that
\be
|0_S\rangle : \qquad \Phi=\Omega=0 ~.
\ee
The vacuum expectation value of the energy-momentum tensor then is
\begin{align}
\langle 0_S|T_{UU}|0_S \rangle &= -\frac1{48\pi U^2} ~,  \\
\langle 0_S|T_{VV}|0_S \rangle &= -\frac1{48\pi V^2}~, \nn \\
\langle 0_S|T_{UV}|0_S \rangle &= -\frac{H^2}{12\pi(H^2UV-1)^2} ~, \nn
\end{align}
which indeed agrees with the known results obtained by a conformal transformation \cite{Candelas:1978gf} or by direct evaluation of mode sums \cite{Birrell:1982ix}. As for the Rindler vacuum, the static vacuum in de Sitter space is singular on both the past and future horizons. It is therefore not a physically reasonable state.

By instead insisting that the energy-momentum tensor be regular at both the future and past horizon, we uniquely single out the Bunch-Davies vacuum
\be
|0_{BD} \rangle : \qquad \Phi=0 ~,\quad \Omega = \frac{H^2}{24\pi} ~.
\ee 
For this state, the expectation value of the energy-momentum tensor is given by
\be
\langle 0_{BD}|T_{\mu\nu}|0_{BD} \rangle = \frac{H^2}{24\pi} g_{\mu\nu} ~,
\ee
in agreement with standard results \cite{Birrell:1982ix}. Even though all free-falling observers see a thermal spectrum of particles \cite{Gibbons:1977}, the incoming and outgoing thermal flux of particles cancel, while the energy density parameter is indeed the appropriate one for a thermal distribution of particles at the de Sitter temperature $T_{dS}=H/2\pi$. In this sense, the Bunch-Davies state is a careful equilibrium of fluxes, just as the Hartle-Hawking state is for black holes. Indeed, in two dimensions the Bunch-Davies vacuum is conformally related to the Hartle-Hawking state for black holes and the Minkowski vacuum in flat space. Furthermore, because the Bunch-Davies state preserves all de Sitter symmetries, since it is proportional to the metric, the energy-momentum tensor does not result in backreaction and just renormalizes the bare cosmological constant.

Finally, let us break the symmetry between incoming and outgoing flux. To construct the Unruh-de Sitter vacuum, we would like to define the incoming sector to be in the Bunch-Davies vacuum and select the static vacuum for the outgoing sector. The energy density in this Unruh-de Sitter vacuum should therefore agree with the incoming energy density $\langle T_{UU} \rangle$ of the Bunch-Davies vacuum and the outgoing energy density $\langle T_{VV} \rangle$ of the static vacuum. To achieve this we need a net amount of positive energy incoming flux, corresponding to a negative value of $\Phi$, and an energy density contribution $\Omega$ that is half that of the Bunch-Davies state
\be \label{eq:Unruh1}
|0_{U} \rangle : \qquad \Phi = -\frac{H^2}{48\pi}~,\quad \Omega = \frac{H^2}{48\pi} ~.
\ee 
The energy-momentum tensor in the Unruh-de Sitter state then indeed equals
\begin{align} \label{eq:Unruh2}
\langle 0_U|T_{UU}|0_U \rangle &= 0 ~,  \\
\langle 0_U|T_{VV}|0_U \rangle &= -\frac1{48\pi V^2}~, \nn \\
\langle 0_U|T_{UV}|0_U \rangle &= -\frac{H^2}{12\pi(H^2UV-1)^2} ~, \nn
\end{align}
which is only singular at the past horizon, as anticipated. Again, the same energy-momentum tensor can also be obtained by a conformal transformation from the Unruh state for the two-dimensional Schwarzschild black hole or by explicitly evaluating the relevant mode sums. 

Let us make a few comments regarding the form of the energy-momentum tensor in the Unruh-de Sitter state. First, notice that the properties of our Unruh-de Sitter state are different from those of the Unruh state for Schwarzschild-de Sitter black holes as constructed in \cite{Markovic:1991ua}. Indeed, taking the limit of vanishing black hole mass, that state reduces to the Bunch-Davies state. Second, in the state that we constructed, an observer at $r=0$ observes only a fixed amount of incoming flux. By removing the outgoing flux, as compared to the Bunch-Davies state, this has resulted in a negative energy density along a constant $V$ slice that violates the null energy condition and is singular at the past horizon. 
In fact, just imposing regularity at the future horizon identifies a one-parameter family of states with the flux given by $\Phi=\Omega-\frac{H^2}{24\pi}$. As a consequence, all these states are well-defined on the entire planar patch and can potentially be used as physically acceptable states in this region. Assuming the static vacuum corresponds to the lowest energy state, the flux parameter is then bounded from below as $\Phi \geq -\frac{H^2}{48\pi}$. This is similar to what happens in the black hole case. There, the Unruh state is constructed by just considering outgoing flux, implying the Hartle-Hawking vacuum for the outgoing sector and the empty Boulware vacuum for the incoming sector. As a result, an asymptotic observer only measures radiation coming from the past horizon, where the energy density actually becomes singular, and the resulting energy-momentum tensor violates the null energy condition \cite{Christensen:1977jc}, allowing the black hole horizon to shrink.

Finally, since the Unruh-de Sitter state is well-defined on the planar patch, it is of interest to consider the energy-momentum tensor, transformed to planar coordinates:
\begin{align} \label{eq:Unruh3}
\langle 0_U|T_{\eta\eta}|0_U \rangle &= -\frac1{24\pi\eta^2} - \frac{1}{48\pi(\eta-\rho)^2} ~,  \\
\langle 0_U|T_{\rho\rho}|0_U \rangle &= +\frac1{24\pi\eta^2} - \frac{1}{48\pi(\eta-\rho)^2}  ~, \nn \\
\langle 0_U|T_{\eta \rho}|0_U \rangle &= \frac{1}{48\pi(\eta-\rho)^2} ~. \nn
\end{align}
In these coordinates, the singularity at the past horizon ($V=H^{-2}(\rho-\eta)^{-1}=0$) is no longer manifest.\footnote{When computing a scalar quantity such as the null energy $T_{\mu\nu}k^\mu k^\nu$, where $k^\mu$ is a null vector, one of course recovers the same singularity as in lightcone coordinates.} As mentioned before, this implies that the Unruh-de Sitter state is an acceptable state on the planar patch. Naively one might have expected any departure from the Bunch-Davies vacuum to redshift away exponentially fast, but by transforming the time-time component to the standard planar time coordinate $\tau$, given by $\partial\eta/\partial \tau=-H\eta$, we discover a continuous source of (incoming) positive energy, which is a corollary of the fact that the energy-momentum tensor in static coordinates is stationary. We would like to stress again that this net flux is absent in the Bunch-Davies state. This can be attributed to the fact that a static observer in the Bunch-Davies states measures equal amounts of incoming and outgoing flux; the Bunch-Davies state is therefore appropriate for describing thermal equilibrium.  Furthermore, in the limit $(-\rho/\eta) \gg 1$ (corresponding to superhorizon scales), the above expressions reduce to the energy-momentum tensor in the Bunch-Davies state. At large superhorizon distances the de Sitter symmetries are restored. The relevant subhorizon contributions are finite, very small, and break some of the de Sitter symmetries (only rotational invariance and time translations are preserved). Of course, these exact results are strictly two-dimensional, so let us now see if and how this result can be generalized to four dimensions. 

\subsection{Generalization to four dimensions}

The method we used to construct the energy-momentum tensor in two dimensions in principle generalizes without much difficulty to four dimensions. The main difference, besides the absence of the conformal anomaly, is that the additional components of the energy-momentum tensor imply more free parameters, unless additional constraints are imposed. By constraining to emission in the s-wave sector we expect the difference between the two and four-dimensional case to be captured by a graybody factor. As we will argue more precisely below, close to the horizon this ensures that the s-wave energy-momentum tensor in four dimensions is effectively two-dimensional. 

Consider a massless scalar field in four-dimensional de Sitter space in static coordinates. The line element is
\be
ds^2 = -\left(1-H^2r^2\right)dt^2 + (1-H^2r^2)^{-1}dr^2  + r^2\left(d\theta^2+\sin^2\theta~ d\phi^2\right)~,
\ee
and the action for a minimally-coupled scalar field is
\be
S = -\frac12\int d^4x\sqrt{-g}~\partial_\mu\varphi\partial^\mu\varphi ~.
\ee
If we restrict to the s-wave sector, $\varphi \equiv \varphi(t,r)$, we can integrate over the two-sphere to obtain an effective two-dimensional action:
\be
S = -2\pi \int d^2x\sqrt{-g_2} ~r^2 \partial_\mu\varphi\partial^\mu\varphi ~.
\ee
Here $g_2$ is the determinant of the metric of two-dimensional de Sitter space. We can go to a canonically normalized field by rescaling:
\be
\varphi \to \tilde\varphi = \frac{1}{\sqrt{4\pi r^2}} \varphi \label{eq:canonical}
\ee
so that the action becomes
\be
S = \int d^2x\sqrt{-g_2}\left(-\frac12\partial_\mu\tilde\varphi\partial^\mu\tilde\varphi - V_{\rm eff}(\tilde\varphi) \right) ~,
\ee
where
\be
V_{\rm eff}(\tilde\varphi) = \frac{(1-H^2r^2)}{2r^2}\left(1 - r\partial_r \right) \tilde \varphi^2 ~.
\ee
We see that the difference between the action of a scalar field in two dimensions and that of the s-wave sector in four dimensions is captured by the effective potential $V_{\rm eff}(\tilde\varphi)$, which vanishes at the horizon $r=1/H$. The radial potential modifies the propagation of modes away from the horizon to the center of de Sitter space, which can effectively be described by graybody factors.

In general, under the assumptions that $\langle T_{\mu\nu} \rangle = \langle T_{\mu\nu}(r) \rangle$ and that the only non-vanishing off-diagonal component is $\langle T_{tr} \rangle$, conservation of the energy-momentum tensor
\be
\nabla_\mu \langle T^{\mu\nu} \rangle = 0 ~,
\ee
leads to the following differential equations
\begin{align} \label{eq:fourdimstress}
\partial_r \langle T^{tr} \rangle &= -\frac{2 \left(2 H^2 r^2-1\right)}{r \left(H^2 r^2-1\right)} \langle T^{tr} \rangle ~, \\
\partial_r \langle T^{rr} \rangle &= -\frac2{r}\langle T^{rr} \rangle + 2r \langle T^{\theta\theta} \rangle - H^2r \langle T^\mu_{\,\,\,\mu} \rangle ~, \nn \\
\langle T^{\phi\phi} \rangle &= \sin^{-2}\theta \, \langle T^{\theta\theta} \rangle \nn~.
\end{align}
An important difference with two dimensions, for which a massless scalar field is automatically conformally invariant, is that a minimally-coupled scalar field in four dimensions is not conformally invariant. Consequently, the trace $\langle T^\mu_{\,\,\,\mu} \rangle$ is not fixed by the conformal anomaly and does not introduce additional constraints. 

The most general solution to \eqref{eq:fourdimstress} is given by
\begin{align}
\langle T_{tt} \rangle &=\frac1{r^2} \left( \Delta - H^2\int_{1/H}^r dr'~ r'^3 \langle T^\mu_{\,\,\,\mu} \rangle + \Theta(r) +2(1-H^2r^2) \langle T_{\theta\theta} \rangle  - r^2(1-H^2r^2) \langle T^\mu_{\,\,\,\mu} \rangle \right) ~,  \\
\langle T_{rr} \rangle &= \frac1{r^2(1-H^2r^2)^2} \left( \Delta - H^2\int_{1/H}^r dr'~ r'^3 \langle T^\mu_{\,\,\,\mu} \rangle + \Theta(r) \right) ~, \nn \\
\langle T_{tr} \rangle &= -\frac{\Phi}{r^2(1-H^2r^2)} ~, \nn 
\end{align}
with $\Phi,\Delta$ constants and
\be
\Theta(r) \equiv 2\int_{1/H}^r dr'~\frac{\langle T_{\theta\theta} \rangle}{r'} ~.
\ee
As before, we see that a non-vanishing $\Phi$ introduces a net flux and breaks the symmetry between incoming and outgoing modes. Also notice that the $\langle T_{tr} \rangle$ component has an apparent singularity at $r=0$ and at $r=1/H$. The singularity at the origin is the four-dimensional generalization of the two-dimensional orbifold singularity, arising due to the vanishing area of the two-sphere. Physically, this implies we need to introduce a source for any non-vanishing net flux $\Phi$. In the Unruh state this source absorbs the incoming positive energy flux from the past horizon. One could resolve this singularity at the origin by assuming a finite size for the physical source. The energy-momentum tensor above can be considerably simplified by assuming the angular components are proportional to the metric and the trace is constant. In that case it is straightforward to see that the Bunch-Davies state corresponds to the parameter choice $\Delta=\Phi=0$, giving $\langle T_{\mu\nu} \rangle \propto g_{\mu\nu}$. Similar to the two-dimensional case we can then again select a one-parameter family of states that contain a net energy flux, but that are nevertheless well-defined on the planar patch by imposing regularity at the future horizon.

Independent of these additional assumptions on the angular components and trace of the stress tensor, we have already seen that the action for the four-dimensional massless minimally coupled scalar in the s-wave sector effectively reduces, at the de Sitter horizon, to the action of a massless minimally coupled scalar field in two dimensions. This implies that close to the horizon the classical energy-momentum tensors in the time and radial directions should be related as follows
\be \label{eq:4d2dmap}
T_{ij}^{(D = 4)} = \frac{H^2}{4\pi} T_{ij}^{(D = 2)} ~,
\ee
where the constant of proportionality follows from the canonically rescaled field \eqref{eq:canonical} evaluated at the horizon, $r = 1/H$. Considering the Bunch-Davies state and comparing the vacuum expectation values of the two- and four-dimensional energy-momentum tensor \cite{Bunch:1978yq}, one finds that the constant of proportionality in \eqref{eq:4d2dmap} is modified and in fact depends on details of the limit to the massless minimally coupled scalar field \cite{Kirsten:1993ug}.

For now ignoring these effects, for the Unruh-de Sitter state one finds using \eqref{eq:Unruh2} that
\begin{align} \label{eq:UnruhStress}
\langle 0_U| T_{UU} |0_U\rangle &= 0 ~, \\
\langle 0_U| T_{VV} |0_U\rangle &= -\frac{H^2}{192\pi^2 V^2}  ~, \nn \\
\langle 0_U| T_{UV} |0_U\rangle &= -\frac{H^4}{48\pi^2} ~, \nn
\end{align}
which has been evaluated at the future horizon ($U=0$). As before, we see that the energy-momentum tensor in the  Unruh-de Sitter state is singular at the past horizon and breaks de Sitter symmetries. It just preserves $SO(3)\times R$, corresponding to rotations and a time translation. The state is inhomogeneous by introducing a continuous source of incoming positive energy (Gibbons-Hawking) flux. As a result one would expect a nontrivial backreaction effect on the initial de Sitter geometry. Also note that, as in the two-dimensional example, the null energy density of the Unruh-de Sitter state is negative. In the class of spherically symmetric, regular, states the Bunch-Davies vacuum now seems special, with incoming and outgoing fluxes precisely cancelling each other. Only for this special (symmetric) choice the energy-momentum tensor is proportional to the de Sitter metric, preserving the de Sitter isometries. This state is therefore appropriate for describing thermal equilibrium but, as we discuss in section \ref{sec:backreaction}, the Unruh-de Sitter state might be a reasonable choice for inflationary spacetimes. 

Now that we have analyzed the vacuum expectation value of the four-dimensional s-wave energy-momentum tensor in the Unruh-de Sitter state close to the horizon, let us next turn to an estimate of the backreaction it should induce on the de Sitter geometry. 

\section{Backreaction and de Sitter evolution} \label{sec:backreaction}

Before describing our analysis of backreaction in the Unruh-de Sitter state, let us once more compare the situation to the black hole case. An important difference is that for black holes one can rely on a globally conserved Hamiltonian, which makes estimating the backreaction effect rather straightforward. In the Unruh state, there is an outgoing flux of positive energy; 
the conservation of energy then immediately implies that the mass of the black hole must decrease. 
In de Sitter space there is no globally conserved energy, so one has to use the local semi-classical Einstein equations. We will estimate the backreaction on the geometry by assuming the Unruh-de Sitter state for a minimally coupled scalar field and inserting the renormalized vacuum expectation value into the semi-classical Einstein equation
\be
G_{\mu\nu} + \Lambda \, g_{\mu\nu}= 8 \pi G_N \, \langle T_{\mu\nu}^{\rm ren} \rangle  ~,
\ee
where $\Lambda$ is the physical cosmological constant and
\be
\langle T_{\mu\nu}^{\rm ren} \rangle \equiv \langle 0_U | T_{\mu\nu} | 0_U \rangle - \langle 0_{BD} | T_{\mu\nu} | 0_{BD} \rangle ~.
\ee
The semi-classical Einstein equation is presumably valid as long as the right-hand side is small. We will assume a semi-classical de Sitter space ($\sqrt{\Lambda} \sim H\ll M_p$) as our initial background, in order to reliably calculate the right-hand side of the equation and then estimate how the small energy-momentum tensor correction affects the geometry on the typical Hubble timescale of the initial de Sitter geometry. To get a first indication of the magnitude of the backreaction effect and confirm the perturbative nature of the correction, one can compare the Unruh-de Sitter energy-momentum tensor expectation value to the classical, cosmological constant, source. This already tells us that if there is a non-trivial effect on the geometry, it will be suppressed as $H^2/M_p^2$. This is again analogous to the situation for evaporating black holes. There $\langle T_{\mu\nu} \rangle$ is calculated for a static Schwarzschild spacetime, even though the outgoing Hawking flux in the Unruh state renders the geometry only approximately static on the typical timescale set by the Schwarzschild geometry. Notice that even though the black hole Unruh state seems to preserve time-translation invariance, by imposing (global) energy conservation one is forced to conclude that the black hole geometry slowly decays. The latter can be understood as related to the singularity at the past horizon in the Unruh state, which should be replaced by collapsing matter, explicitly identifying an initial time and breaking the time-translation symmetry. Similarly, we would like to interpret the breaking of the time-translation symmetry in the Unruh-de Sitter state as due to the presence of the past horizon singularity, forcing one to pick an initial time, and correspondingly, an initial Hubble parameter. 

\begin{figure}[ht]
    \centering
    \includegraphics[scale=1.0]{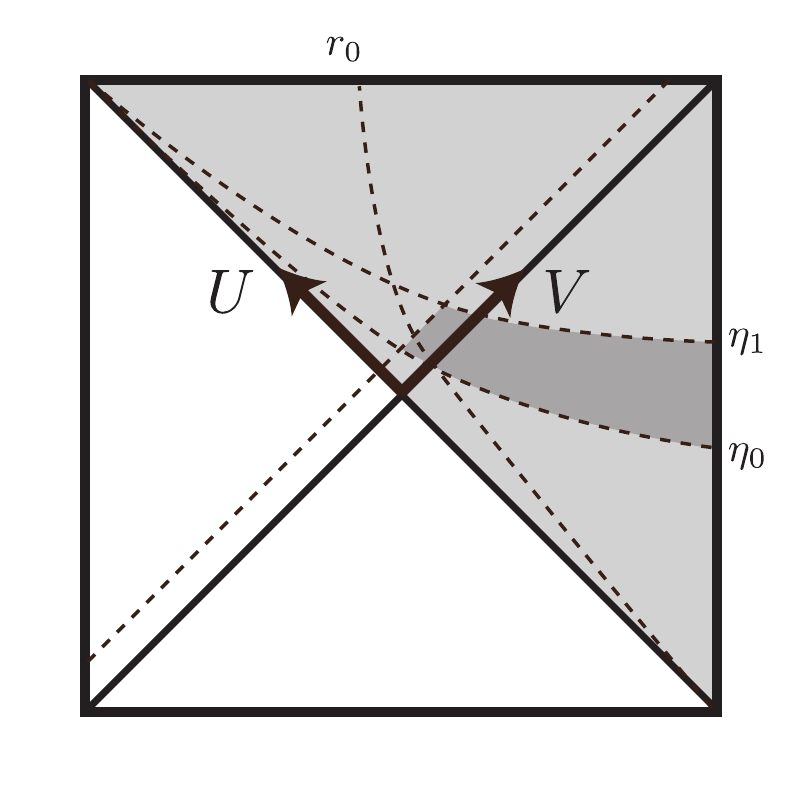}
    \caption{Illustration of the finite region (dark shaded) in the de Sitter Penrose diagram where we can analyze the approximately homogeneous backreaction effect due to the stress-energy in the Unruh state. Here $|r_0/\eta_0| \gtrsim 1$ and $\eta_1$ corresponds to the time where the effect becomes order one.}
    \label{dSbackreaction}
\end{figure}

It is important to emphasize that we have derived the four-dimensional energy-momen\-tum tensor in the Unruh-de Sitter state only in the vicinity of the horizon. Nevertheless, to estimate the effects on the de Sitter geometry we will use a homogeneous and isotropic ansatz. This is, strictly speaking, incorrect: the expectation value of the energy-momentum tensor has not only an explicit radial dependence breaking homogeneity, but also an off-diagonal term. To estimate the backreaction we will take the near-horizon result and use it everywhere in a volume of roughly order one Hubble radius, assuming that on large superhorizon scales the energy-momentum tensor reproduces the energy-momentum tensor in the Bunch-Davies state. In the two-dimensional case, this is indeed exactly what happens \eqref{eq:Unruh3} and although it seems reasonable to assume this remains true in four dimensions, it would be worthwhile to verify this more explicitly in future work. The situation is sketched in Figure \ref{dSbackreaction}, where $\eta_0$ and $r_0$ identify the initial time and finite co-moving size of a potential-dominated homogeneous region of spacetime. This patch is inflating at a Hubble rate $H_0\equiv H(\eta_0)$. The shaded area corresponds to the region where the backreaction effect can be estimated.

Let us analyze the backreaction by starting with a homogeneous and isotropic Friedmann-Lema\^itre-Robertson-Walker line element
\be
ds^2 = -d\tau^2 + a^2(\tau)\left( d\rho^2 + \rho^2(d\theta^2 + \sin^2\theta ~d\phi^2)\right) ~,
\ee
where $\tau$ is now the planar time coordinate and $a(\tau)$ the scale factor. Combining the time-time and spatial-spatial components of the Einstein equations, the evolution of the Hubble parameter is given by
\be \label{eq:evolutionequation}
\dot H = -\frac{4\pi G_N}{a(t)^2}\left(\langle T_{\eta\eta} \rangle +\langle T_{rr} \rangle \right) ~,
\ee
where the dot denotes a time derivative with respect to $\tau$ and $\langle T_{\eta\eta} \rangle$ is the time-time component of the energy-momentum tensor using conformal time $\partial\eta/\partial \tau=-H\eta$.

We will now use the value of the near-horizon energy-momentum tensor as a proxy for the  energy-momentum everywhere in a Hubble-sized region. Again, in two dimensions this is an excellent approximation and we will assume this remains true in four dimensions.  We then find (using \eqref{eq:UnruhStress})
\begin{align}
\langle 0_U| T_{\eta\eta} |0_U\rangle &= -\frac{3H^2}{256\pi^2\eta^2} ~, \\
\langle 0_U| T_{\rho\rho} |0_U\rangle &= \frac{7H^2}{768\pi^2\eta^2}  ~, \nn \\
\langle 0_U| T_{\eta \rho} |0_U\rangle &= \frac{H^2}{768\pi^2\eta^2}  ~. \nn
\end{align}
Furthermore, just as in the black hole case we will choose to ignore the off-diagonal component, since it is not expected to affect the (local) Hubble parameter to leading order. In a more complete treatment it would certainly be interesting to include all components and verify this explicitly.  

Plugging in the small and approximately homogeneous and isotropic energy-momentum tensor correction, the evolution equation for the Hubble parameter is (with $N \equiv \log{a}$ the number of e-folds and $M_p^{-2} = 8\pi G_N$ the reduced Planck mass) 
\be
\dot H \approx \frac{H^4}{768\pi^2 M_p^2} \quad \rightarrow \quad \frac{1}{H}\frac{dH}{dN} \approx \frac{1}{768 \pi^2} \frac{H^2}{M_p^2}
\ee
implying that the relative magnitude of the backreaction effect per e-fold is suppressed as $(H/M_p)^2$, as we already anticipated. The inverse can then be associated to a bound on the lifetime of de Sitter space that is of the order of the de Sitter entropy
\be
N_{\mathrm{max}} = H \, \tau \sim M_p^2/H^2 \propto S_{dS}~.
\ee
As observed earlier, the energy-momentum tensor in the Unruh-de Sitter state violates the null energy condition. By again referring to the analogy with black holes this should not come as a complete surprise: we know that black holes evaporate in the Unruh state and, by Hawking's area theorem, a decrease in the horizon area is only possible by violating the null energy condition locally, at the scale of the horizon. It would be interesting to study more general energy conditions such as the averaged null energy condition or the quantum null energy condition in this context. For now we conclude that the presence of a (small and negative) vacuum energy and pressure correction in the Unruh-de Sitter state effectively causes the Hubble radius to {\em decrease}. The evolution can be followed as long as the effects remain in the semi-classical regime, until it reaches a presumably highly curved, and ultimately singular regime, beyond an effective field theory description. 

Let us make a few final comments. To obtain an estimate of the effect of backreaction on the background geometry, we had to make some assumptions regarding the form of the energy-momentum tensor in four dimensions. Based on the two-dimensional result, we assumed the difference between the Unruh state and the Bunch-Davies vacuum to be confined to a Hubble-sized region. In addition, we also assumed the near-horizon value of the energy-momentum tensor to provide a reasonable estimate of the true energy-momentum in a Hubble-sized region and we ignored the off-diagonal component of the energy-momentum tensor. It is important to stress however that, despite the subtleties in trying to track the evolution of the backreacted geometry, the conclusion that this Hubble-size region is unstable in the Unruh-de Sitter state seems to be unavoidable. Contrary to a (slightly perturbed) exactly homogeneous and isotropic energy-momentum tensor the continuous flux of stress energy in the Unruh-de Sitter state does not redshift away to reduce to the Bunch-Davies state, and as a consequence it breaks the de Sitter isometries and will non-trivially backreact. As for black holes in the Unruh state, the Unruh-de Sitter state contains negative outgoing null energy (given by $\langle T_{VV} \rangle$), because of the removal of outgoing flux as compared to the Bunch-Davies vacuum. This small amount of negative outgoing null energy violates the null energy condition and causes the de Sitter horizon to shrink. 

This somewhat surprising conclusion is further confirmed by the physics. In static coordinates we have seen that the Unruh-de Sitter state contains a continuous source of positive energy incoming flux. As is well known, adding energy to the static region indeed shrinks the cosmological horizon, which has for instance been used to compute corrections to the Bunch-Davies vacuum consistent with the obtained result \cite{Greene:2005wk}. This incoming positive energy flux is equivalent to an outgoing negative energy flux, passing through the horizon. This negative energy outgoing flux, in an analysis using planar (or Painlev\'e) coordinates, is violating the null energy condition and produces a small, but growing, correction to the Hubble parameter. Clearly though, in future work it would be of interest to go beyond the approximations we made here, and track the evolution of the instability more precisely using cosmological perturbation theory.

\subsection{Consistency and interpretation}

Although exact global de Sitter space is a very interesting spacetime, the typical cases of physical interest require only a geometry that is approximately de Sitter space in a few Hubble-size regions of space and time. Examples include i) the interior of a bubble of false vacuum and ii) a patch of space dominated by vacuum energy leading to inflation. In both these cases, the local geometry is essentially that of de Sitter space. However, certain global features of global de Sitter space -- in particular, the existence of past horizons -- are not necessarily present. Therefore the existence of a $\langle T_{\mu\nu} \rangle$ that is singular at the past horizon is not immediately problematic. Again this is analogous to physical black holes that form from collapsing matter. In these scenarios the Unruh-de Sitter state therefore seems to be a plausible candidate for an initial state. More generally, any linear combination of the Bunch-Davies state and the Unruh-de Sitter state would also be acceptable. As a first step, in this work we constructed the Unruh-de Sitter state and presented a first analysis of the consequences for the subsequent evolution of de Sitter space.

If one were interested in using the Unruh-de Sitter state for cosmological purposes, one might worry that the introduction of an inhomogeneous initial state might be in direct conflict with observation. Inflationary states should produce a large, flat and extremely homogeneous observable universe. In fact, selecting the Unruh-de Sitter vacuum does not alter this expectation. The state picks out a preferred point, which can be regarded as the center of a finite potential-dominated region containing a homogeneous scalar field. This patch then exponentially expands and can accommodate the entire observable universe after roughly 60 e-folds of inflation. The inherent instability of the Unruh-de Sitter state becomes apparent only after $S_{dS}$ e-folds, which is much larger than 60 e-folds for reasonable values of the inflationary Hubble parameter. Additionally, in slow-roll models of inflation the `backreaction' effect due to classical evolution is always much larger as long as one avoids the eternal inflation regime, forcing $\epsilon \gg H^2/M_p^2$.
We therefore expect a large suppression of these corrections to inflationary correlation functions. In other words, the effects due to the inhomogeneous nature of the Unruh-de Sitter state can safely be ignored in the context of $60$ e-folds of slow-roll inflation. This possibly also relates to the following observation \cite{Greene:2005wk}: even though the static vacuum identifies a special observer, breaking homogeneity, all other free-falling observers will have a hard time to distinguish the static vacuum from the Bunch-Davies state. Formally one can identify the Bunch-Davies vacuum with the infinite boost light-like limit of the static vacuum. In other words, the properties of the Unruh-de Sitter state, as long as the number of e-folds is much smaller than the de Sitter entropy, are expected to be indistinguishable from the Bunch-Davies vacuum. We hope to explicitly confirm these expectations in future work. 

Another point of concern with the Unruh-de Sitter state is how the decrease of the de Sitter horizon can be consistent with thermodynamics. De Sitter thermodynamics and even the meaning of de Sitter entropy is a subtle subject \cite{Gibbons:1977,Kastor:1993mj,Spradlin:2001pw,Witten:2001kn,Banks:2000fe,Parikh:2004wh,Parikh:2004ux,Parikh:2008iu,Bhattacharya:2018ltm,Dinsmore:2019elr,Qiu:2019qgp}. Naively, the decrease in the horizon radius seems to violate the second law:
\be
\dot S_{horizon} = \frac{\dot A}{4 \, G_N} = -\frac{2\pi}{H^3 G_N}\dot H  < 0 ~.
\ee
Of course, for black holes we know how this works. By taking into account the entropy of the emitted radiation, the total generalized entropy still increases. This suggests that in (a Hubble-size region of) de Sitter space, something similar could happen. To calculate the total entropy, we will make some assumptions. First, we will work in planar coordinates and consider the entropy change in a fixed proper volume, $V$. Also, as before, we will regard the vacuum energy to be homogeneous and isotropic. At fixed volume, the entropy change is given by
\be
dS_{\rm vacuum} = \frac{dE}{T} ~,
\ee
By using the continuity equation
\be
\dot \varepsilon = -3H(\varepsilon+p) ~,
\ee
where $\rho$ is the energy density and $p$ the pressure we can express this as
\be
\dot S_{\rm vacuum} = -\frac{3H}{T}(\varepsilon+p)V ~,
\ee
where the dot denotes a derivative with respect to planar time, $t$. Any contribution from a cosmological constant ($\varepsilon=-p$) drops out, as expected. By combining this with the result for the change in horizon area and by using the Friedmann equations, the total entropy change can be written as
\be \label{eq:totentropy}
\dot S_{\rm total} = \dot S_{\rm horizon}+ \dot S_{\rm vacuum} =  \left(\frac{8\pi^2}{H^3} - \frac{3HV}{T} \right)(\varepsilon+p) ~.
\ee
If we make the further assumption that the vacuum temperature is uniform and given by the de Sitter temperature, we find that both entropy contributions in a fixed Hubble volume $V = \frac{4\pi}{3H^3}$ cancel exactly such that
\be
\dot S_{\rm total} = 0 ~,
\ee
which saturates the second law. This behavior was already observed in \cite{Davies:1987ti}, indicating that a decrease of the horizon area is not necessarily in conflict with the second law when applied to a single Hubble-sized volume.

That the generalized entropy vanishes in a Hubble volume is also supported by Bousso's $N$-bound \cite{Bousso:2000nf}, which states roughly that the entropy in a causal diamond in a spacetime with a positive cosmological constant is bounded from above by the de Sitter entropy. Because the static patch region is the largest causal diamond in de Sitter, it must be that $S_{\rm total} \leq S_{dS}$. At the same time, if we assume the initial entropy to be given by the de Sitter entropy, then the generalized second law implies $S_{\rm total}\geq S_{dS}$. The only way these two bounds can be consistent with each other is if the total entropy in a static patch region remains constant.

For volumes larger than the proper volume of a static patch, the $N$-bound no longer applies and the vacuum entropy term in \eqref{eq:totentropy} dominates over the horizon contribution. We then have
\be
\dot S_{\rm total} > 0 ~,
\ee
provided that $(\varepsilon+p) < 0$. Notably this is the case in the Unruh-de Sitter state. We see that, if these back-of-the-envelope calculations and their underlying assumptions are valid, then a decrease in horizon area is not necessarily in conflict with the second law of thermodynamics.

\section{Summary and Discussion} \label{sec:conclusions}

In this paper, we have considered a massless minimally coupled scalar field in a de Sitter background in a novel state -- the Unruh-de Sitter state -- which is the de Sitter counterpart of the black hole Unruh state. We computed the vacuum expectation value of the energy-momentum tensor in the Bunch-Davies, static, and Unruh-de Sitter vacua and related the two-dimensional energy-momentum tensor to the near-horizon energy-momentum tensor in four dimensions in the s-wave sector. We then argued that the Unruh-de Sitter state, which is only singular at the past horizon, might be a viable and natural alternative to the de Sitter-invariant Bunch-Davies state typically assumed in inflationary applications. Modifications to the standard predictions from $60$ e-folds of slow-roll inflation assuming the Bunch-Davies state are expected to be negligible. However, it does suggest a fundamental bound on the maximum number of e-folds given by the de Sitter entropy for at least a Hubble-sized region, before one enters a strongly curved regime. This could have important consequences in the context of eternal inflation, and related to that, anthropic scenarios involving the string landscape.

Such a bound might also be relevant in the context of the recent (refined) de Sitter swampland conjectures \cite{Obied:2018sgi,Ooguri:2018wrx}. Note however that, first of all, the instability in the Unruh-de Sitter state is much slower than in any of the de Sitter swampland conjectures, removing any potential tension with single field slow-roll inflation or dark energy constraints. Moreover, the instability evolves towards smaller de Sitter radii which, as we have explained, can be understood physically by the presence of a continuous source of negative energy outgoing flux. This is to be contrasted with the de Sitter swampland conjectures, for which the de Sitter radius increases. This different behavior is not in conflict with our results, as the physical system under consideration is rather different; in the de Sitter swampland conjectures it is assumed that a tower of states becomes light and contributes to the de Sitter entropy, whereas we are just relying on an effective analysis of a single massless scalar field in the de Sitter background. Our results therefore seem to be more closely related to the so-called ``quantum break time" of de Sitter space that was recently revisited in \cite{Dvali:2018jhn}, and perhaps to the vacuum state modifications suggested in \cite{Danielsson:2018qpa}.

The physical origin of the backreaction effect is clear. Similar to black holes, one identifies a special static observer who measures a continuous positive energy flux of incoming thermal radiation or, equivalently, an outgoing flux of negative energy. This requires the presence of a physical source at the origin, breaking homogeneity, and results in a backreaction effect that shrinks the horizon ever so slightly. Our conclusion is also supported by the complementary point of view in terms of radiation tunnelling through the de Sitter horizon \cite{Parikh:2002qh}. In an effective field theory description of the tunneling process in de Sitter space one indeed requires the identification of a state that is empty for an observer freely falling through the horizon. As for black holes \cite{Aalsma:2018qwy}, this naturally selects the Unruh state. The asymmetric nature of the state is crucial, since it introduces unbalanced incoming and outgoing fluxes, breaking equilibrium. Indeed, within the class of all isotropic states that are non-singular on the future horizon, the Bunch-Davies state would now seem rather fine-tuned.

By allowing isotropic vacuum states that introduce singularities only at the past horizon, the situation in de Sitter becomes analogous to that for black holes. For initial Hubble parameters far below the Planck scale the backreaction is tiny and measured in terms of the de Sitter entropy. As a consequence we do not expect any important effects in the context of slow-roll inflationary phases, which require only of the order of $10-100$ e-folds. But it would certainly be interesting to show this explicitly in terms of correlation functions. Finally, we speculate that by introducing a family of Unruh-de Sitter states, all producing large, flat, and approximately homogeneous universes after a finite phase of slow-roll inflation, one could alleviate some fine-tuning issues related to the initial conditions of inflation. We hope to report on many of these remaining questions in future work.

\acknowledgments
We would like to thank Manus Visser, Ben Freivogel, and Thomas van Riet for useful discussions. LA would like to thank the University of Wisconsin-Madison for hospitality while part of this work was completed. This work is part of the Delta ITP consortium, a program of the Netherlands Organisation for Scientific Research (NWO) that is funded by the Dutch Ministry of Education, Culture and Science (OCW). This work is also supported in part by the Foundation for Fundamental Research (FOM), which is part of NWO. MP is supported in part by the John Templeton Foundation grant 60253.

\appendix

\section{Energy-momentum tensor in the Rindler vacuum}

In this appendix, we will calculate the expectation value of the energy-momentum tensor of a free scalar field in the Rindler vacuum. We have
\be
\langle T_{\mu \nu} \rangle = \langle \partial_\mu \varphi \partial_\nu \varphi \rangle - \frac{1}{2} \eta_{\mu\nu}\langle\partial_\rho\varphi\partial^\rho\varphi\rangle ~.
\ee
This expression becomes particularly simple in the lightcone coordinates
\be
U = t-x ~, \qquad V = t+x ~,
\ee
where we find
\begin{align}
\langle T_{UU}\rangle &= \langle\left(\partial_U\varphi\right)^2\rangle \\
\langle T_{VV}\rangle &= \langle\left(\partial_V\varphi\right)^2\rangle \nn \\
\langle T_{UV}\rangle &= 0 ~. \nn
\end{align}
To evaluate these expressions, we need to expand the scalar field in mode functions. An expansion in modes that are positive frequency with respect to a Minkowski observer is given by
\be
\hat\varphi = \int_0^\infty \frac{d\omega}{\sqrt{4\pi \omega}} \left[ e^{-i\omega U}~\hat a_\omega + e^{+i\omega U}~\hat a^\dagger_\omega \right] + (U \leftrightarrow V) ~.
\ee
The creation/annihilation operators obey the canonical commutation relations
\be
[\hat a_\omega, \hat a_{\omega'}^\dagger] = \delta(\omega-\omega') ~.
\ee
Similarly, an expansion in modes that are positive frequency with respect to a Rindler observer in the right Rindler wedge is given by \cite{Parentani:1993yz}
\be \label{eq:rindlermodes}
\hat\varphi = \int_0^\infty \frac{d\omega}{\sqrt{4\pi \omega}} \left( (-aU)^{i\omega/a}\hat b_\omega + (-aU)^{-i\omega/a}\hat b^\dagger_\omega \right) + (U \leftrightarrow V) ~,
\ee
where
\be
[\hat b_\omega, \hat b_{\omega'}^\dagger] = \delta(\omega-\omega') ~.
\ee
These two sets of modes are related by a Bogolubov transformation:
\be \label{eq:bogotrafo}
\hat b_\omega = \int_0^\infty d\omega'\left(\alpha_{\omega\omega'}\hat a_{\omega'} + \beta_{\omega\omega'}\hat a^\dagger_{\omega'}\right) ~,
\ee
where
\be
|\beta_\omega|^2 = \frac1{e^{2\pi \omega/a}-1} ~.
\ee
We can now evaluate the expression for the energy-momentum tensor by expanding the scalar field in Rindler modes \eqref{eq:rindlermodes}. We find
\begin{align} \label{eq:stressexpr}
\braket{(\partial_U \varphi)^2} &= \int_0^\infty \frac{d\omega~\omega}{4\pi a^2U^2}\left(1+2\braket{\hat b_\omega^\dagger\hat b_\omega}\right) \\
\braket{(\partial_V \varphi)^2} &= \int_0^\infty \frac{d\omega~\omega}{4\pi a^2V^2}\left(1+2\braket{\hat b_\omega^\dagger\hat b_\omega}\right) \nn
\end{align}
The first term on the right-hand side of \eqref{eq:stressexpr} is UV-divergent and we need to impose some regularization procedure to obtain a finite answer. Because the Rindler vacuum and Minkowski vacuum contain the same UV divergence we can obtain a regular answer by considering the difference between the energy-momentum tensor in the Rindler and Minkowski vacuum. Using the expressions \eqref{eq:stressexpr} we obtain
\begin{align}
\langle 0_R|T_{UU}|0_R\rangle - \langle 0_M| T_{UU}|0_M\rangle &= - \int_0^\infty \frac{d\omega~\omega}{2\pi a^2U^2} \langle 0_M| \hat b_\omega^\dagger\hat b_\omega  | 0_M\rangle ~,\\
\langle 0_R|T_{VV}|0_R\rangle - \langle 0_M| T_{VV}|0_M\rangle &= - \int_0^\infty \frac{d\omega~\omega}{2\pi a^2V^2} \langle 0_M| \hat b_\omega^\dagger\hat b_\omega  | 0_M\rangle ~. \nn
\end{align}
Then, using the expression for the Bogolubov transformation \eqref{eq:bogotrafo} and evaluating the integral, we obtain the result \cite{Parentani:1993yz}
\begin{align}
\langle 0_R|T_{UU}|0_R\rangle - \langle 0_M| T_{UU}|0_M\rangle &=  -\frac1{48\pi U^2} ~,\\
\langle 0_R|T_{VV}|0_R\rangle - \langle 0_M| T_{VV}|0_M\rangle &= -\frac1{48\pi V^2} ~. \nn
\end{align}
Locally, the Rindler vacuum has a negative energy density as compared to the Minkowski vacuum.

\bibliographystyle{JHEP}
\bibliography{refs}

\end{document}